\documentstyle[vsolj01,graphicx,natbib]{article}

\RequirePackage[T1]{fontenc}

\def\cite{\citealt}

\begin{document}

\title{ZTF J185139.81$+$171430.3 = ZTF18abnbzvx: the second white dwarf pulsar?}

\author{Taichi Kato$^1$ and Naoto Kojiguchi$^1$}
\author{$^1$ Department of Astronomy, Kyoto University,
       Sakyo-ku, Kyoto 606-8502, Japan}
\email{tkato@kusastro.kyoto-u.ac.jp}

\begin{abstract}
We found that ZTF J185139.81$+$171430.3 = ZTF18abnbzvx shows
large-amplitude (0.8~mag) coherent variations with
a very short [0.00858995(3)~d = 12.37~min] period using
Public Data Release of Zwicky Transient Facility
observations.  The only known object that shows similar
very short period, large-amplitude and coherent variations
is the unique white dwarf pulsar AR Sco.
The variations in ZTF J185139.81$+$171430.3 may arise from
a mechanism as in AR Sco and this object should deserve attention.
\end{abstract}

   ZTF18abnbzvx is a variable object detected by
the Zwicky Transient Facility (ZTF) project.  The object
was classified as a dwarf nova by the Automatic Learning for the
Rapid Classification of Events (ALeRCE) Alert Broker
\citep{for21ALeRCE}.  The object was listed as a candidate
variable star (ZTF J185139.81$+$171430.3, hereafter ZTF J1851)
having a range of 17.561--18.836 and a period of 0.0086~d
in table 4 of \citet{ofe20ZTFvariables}.
Automatic period detections often give spurious periods 
(either close to 1~d, 0.5~d or very short ones) and
such periods are not usually considered seriously.

\begin{figure*}
  \begin{center}
    \includegraphics[width=16cm]{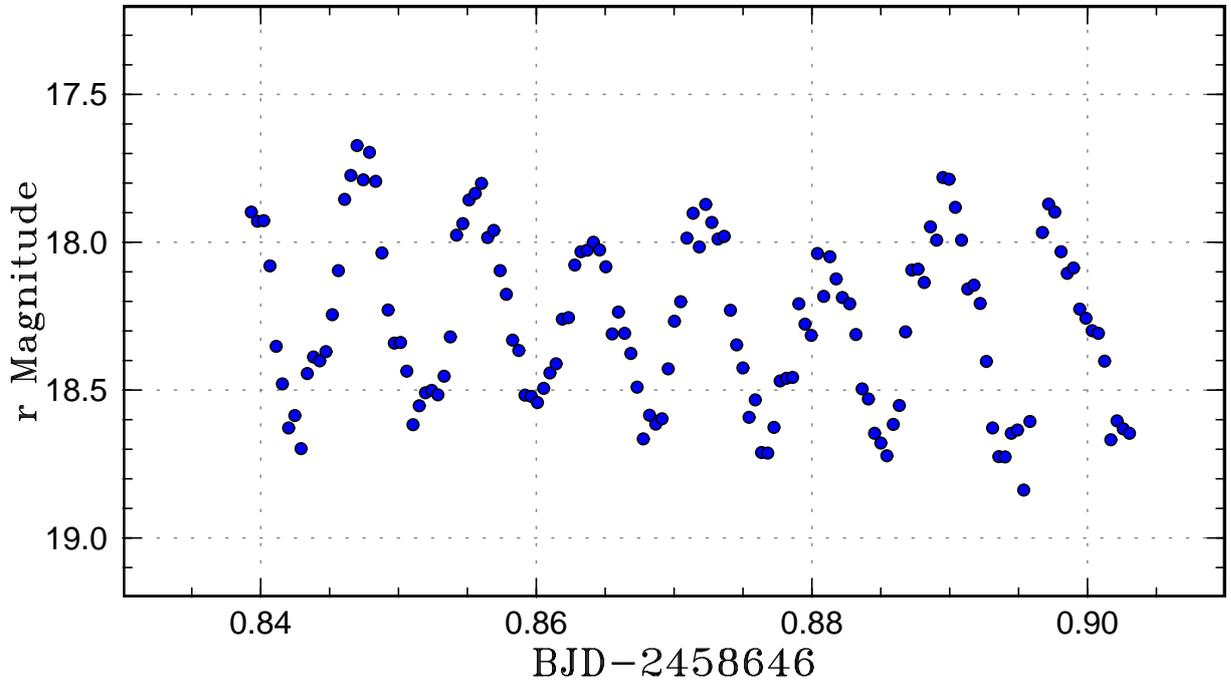}
  \end{center}
  \caption{Short-period variation in ZTF J185139.81$+$171430.3 = ZTF18abnbzvx.}
  \label{fig:short}
\end{figure*}

\begin{figure*}
  \begin{center}
    \includegraphics[width=16cm]{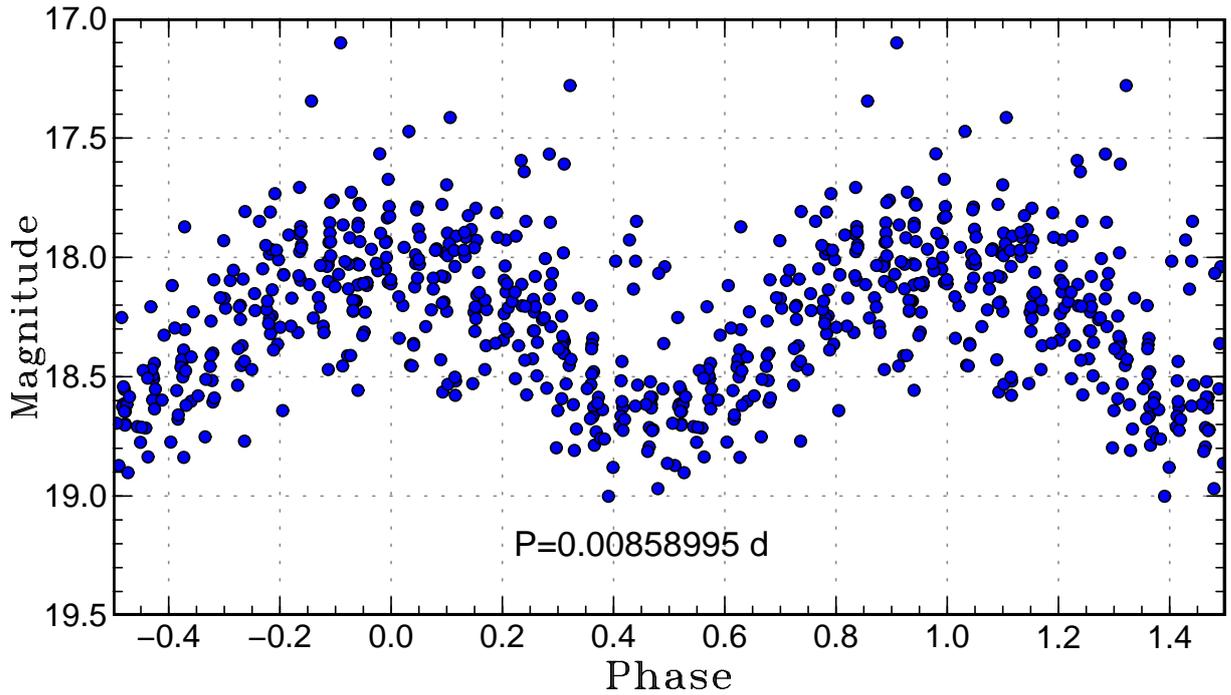}
  \end{center}
  \caption{ZTF $r$ phase-folded light curve of
    ZTF J185139.81$+$171430.3 = ZTF18abnbzvx
    for the segment BJD 2458675--2458746.
    The epoch was chosen as BJD 2458737.3078.}
  \label{fig:phase}
\end{figure*}

   We used Public Data Release 6 of
the Zwicky Transient Facility \citep{ZTF}
observations\footnote{
   The ZTF data can be obtained from IRSA
$<$https://irsa.ipac.caltech.edu/Missions/ztf.html$>$
using the interface
$<$https://irsa.ipac.caltech.edu/docs/program\_interface/ztf\_api.html$>$
or using a wrapper of the above IRSA API
$<$https://github.com/MickaelRigault/ztfquery$>$.
} and found that this object showed large-amplitude
(0.6--1.0~mag) and very short period variations
during a time-resolved run on 2019 June 12 (BJD 2458646;
figure \ref{fig:short}).
We found that the period is in agreement with
the candidate period in \citet{ofe20ZTFvariables}.
We also confirmed that the variations recorded
for BJD 2458675--2458746, the segment with multiple
nightly observations, could be expressed by the same period.
By using the entire data BJD 2458646--2458746,
we obtained a period of 0.00858995(3)~d.
The resultant phased light curve (figure \ref{fig:phase})
shows that that the variations are coherent.

   The variation arising from the orbital variation
(such as a reflection variable)
is very unlikely considering the large amplitude
(0.8~mag) and the very short (12.37~min) period.
The only known class of binaries with comparably short
periods are AM CVn stars (see e.g. \cite{sol10amcvnreview}).
No known AM CVn stars with similar periods show such
large-amplitude orbital modulations.
The small parallax of 0.325(119) mas in \citet{GaiaEDR3}
also does not favor an underluminous binary.

   Stellar pulsations are also unlikely.  The large
amplitude requires radial pulsations, but there is
no known class of radially pulsating variables with 
this short period.

   We propose that this short period may reflect
the spin period of the white dwarf.  The famous
white dwarf pulsar AR Sco \citep{mar16arsco}, which is
a 3.56-hr binary consisting of a white dwarf and an M5 star.
The spin period of the white dwarf in AR Sco is 1.97~min
and amplitudes of the pulses reach 0.5--1.0~mag
in the optical (\cite{mar16arsco}; \cite{sti18arsco}).

\begin{figure*}
  \begin{center}
    \includegraphics[width=16cm]{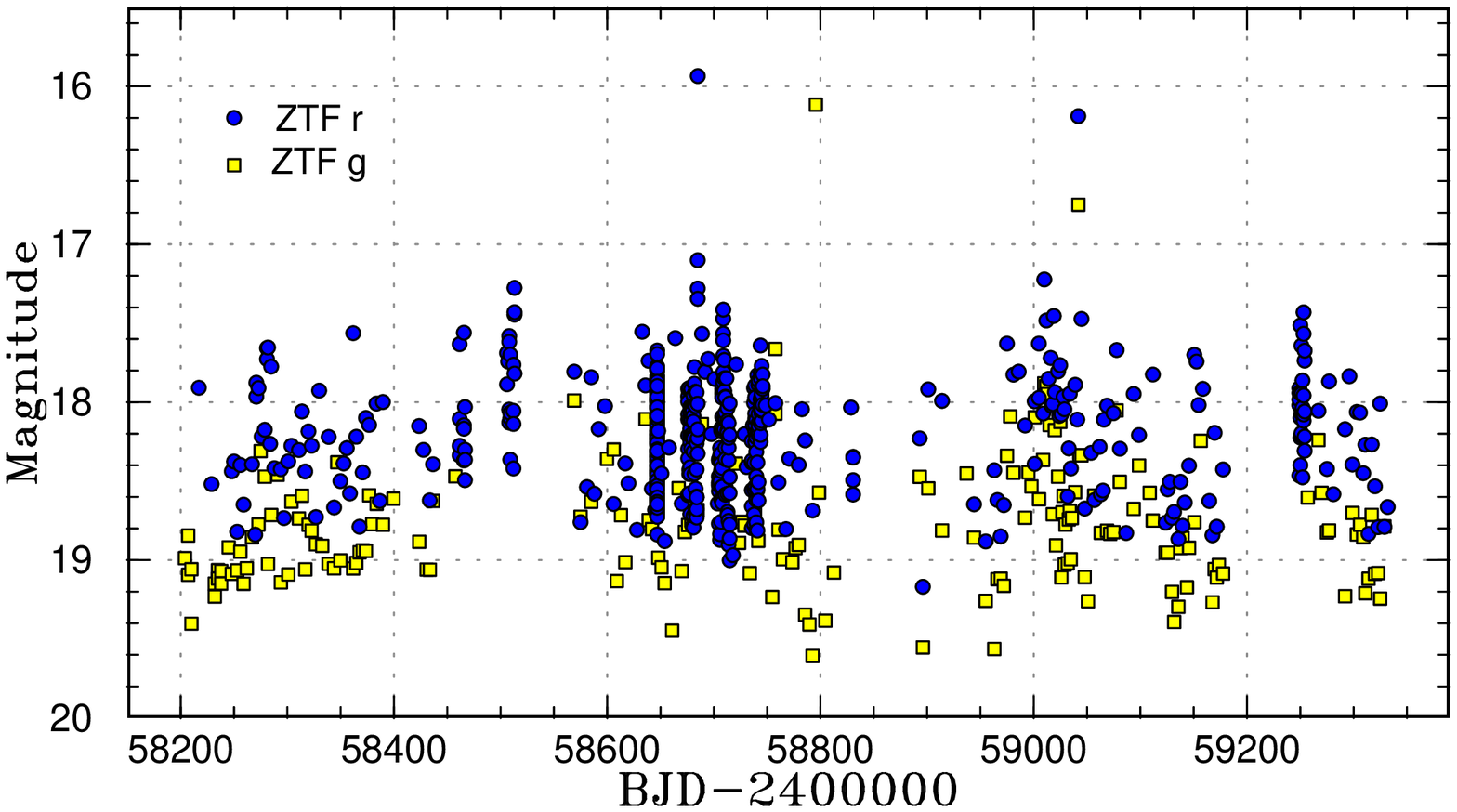}
  \end{center}
  \caption{Long-term light curve of ZTF J185139.81$+$171430.3 = ZTF18abnbzvx
     from ZTF observations.}
  \label{fig:lc}
\end{figure*}

   In contrast to AR Sco, orbital modulations are not
apparent in ZTF J1851 (figure \ref{fig:lc}).
There were possible outbursts with amplitudes of $\sim$2~mag
such as on BJD 2458685 and on BJD 2459042.
The reality of these possible outbursts requires confirmation
since there were only a few observations during these events
A quasi-simultaneous rise in $g$ and $r$ was recorded on
BJD 2459042.  The object returned to the normal brightness
on the subsequent day.  These ``outbursts'', if present,
should have been short-lived as in outbursts in intermediate
polars.  Although further observations are needed to see whether
the short-period coherent variations in ZTF J1851 indeed
arises from a spinning white dwarf, this object would be
a good candidate for a white dwarf pulsar and should deserve attention.

\section*{Acknowledgments}

This work was supported by JSPS KAKENHI Grant Number 21K03616.

Based on observations obtained with the Samuel Oschin 48-inch
Telescope at the Palomar Observatory as part of
the Zwicky Transient Facility project. ZTF is supported by
the National Science Foundation under Grant No. AST-1440341
and a collaboration including Caltech, IPAC, 
the Weizmann Institute for Science, the Oskar Klein Center
at Stockholm University, the University of Maryland,
the University of Washington, Deutsches Elektronen-Synchrotron
and Humboldt University, Los Alamos National Laboratories, 
the TANGO Consortium of Taiwan, the University of 
Wisconsin at Milwaukee, and Lawrence Berkeley National Laboratories.
Operations are conducted by COO, IPAC, and UW.

The ztfquery code was funded by the European Research Council
(ERC) under the European Union's Horizon 2020 research and 
innovation programme (grant agreement n$^{\circ}$759194
-- USNAC, PI: Rigault).

\end{document}